 \documentclass[prb,aps,twocolumn,showpacs]{revtex4} 
\usepackage{graphicx} 
\usepackage{tabularx}
\usepackage{dcolumn} 
\usepackage{color}
\usepackage{bm}
\usepackage{amsmath}
\usepackage{amssymb}

\begin{document} 


\title{ Violation of the orbital depairing limit in a non-unitary state \\
--on the high field phase in the heavy Fermion superconductor UTe$_2$--
}

\author{Kazushige Machida} 
\affiliation{Department of Physics, Ritsumeikan University, 
Kusatsu 525-8577, Japan} 

\date{\today}

\begin{abstract}
A theoretical study is reported on the origin of extremely high upper critical field $\sim$70T observed in 
UTe$_2$ with the transition temperature T$_{\rm c}$=1.6K-2K, far exceeding the conventional orbital depairing limit 
set by the Fermi velocity and T$_{\rm c}$ for a superconductor (SC) in the clean limit.  We investigate possible
violation of the orbital limit in terms of a spin-triplet nonunitary state, which is effectively 
coupled to the underlying magnetization induced by external field. This in turn produces the
reduced internal field by cancelling it via magnetization. We formulate a theory within Ginzburg-Landau framework
to describe this orbital limit violation and analyze experimental data on the upper critical fields
for various field orientations in UTe$_2$. It is pointed out that the orbital limit violation for a spin-triplet SC together with
the Pauli-Clogston limit violation for a spin-singlet SC constitutes a complete and useful framework to examine the
high field physics in superconductors in the clean limit.
 \end{abstract}

 
 
\maketitle 

\section{Introduction}

Much attention has been focused on a recently found heavy Fermion superconductor UTe$_2$
because of a candidate material of a triplet pairing, which is quite rare except for 
superfluid $^3$He~\cite{3he,mizushima} and UPt$_3$~\cite{upt3,ohmi,yo,tsutsumi1}.
They are all characterized by multiple phases due to rich internal degrees of freedom
inherent to a spin-triplet pairing.
UTe$_2$ is known to exhibit remarkable superconducting (SC) properties
in addition to multiple phases in magnetic field (H) and temperature (T)
plane under both ambient and applied pressure~\cite{review}.
In the SC energy gap structure probed by several thermodynamic measurements~\cite{review,metz,kittaka}
 a pair of point nodes is situated along the
$a$-axis in orthorhombic crystal.
The time reversal symmetry is broken in the SC phase detected by the Kerr rotation
experiment~\cite{kaptulnik}.
The scanning tunneling microscopy (STM) experiment suggests that the chiral SC may be realized~\cite{madhavan}.

According to a series of $^{125}$Te NMR experiments~\cite{ishida1,ishida2,ishida3,ishida4,ishida5}, 
the Knight shift (KS) or
the spin susceptibility drops (remains uncharged) along the $b$-axis and $c$-axis (the $a$-axis)
 below the SC transition temperature T$_{{\rm c}}$ at low fields, showing that
 the d-vector points perpendicular
to the $a$-axis. Namely the d-vector has the components along the the $b$ and $c$-axes.
 At the lowest fields  along the $b$-axis the KS decreases, but as $H$ increases
from 5T up to $\sim$12T the KS as a function of $H$ gradually ceases decreasing
to return to the normal value.
This implies that the d-vector changes its direction so as to be perpendicular to the
applied field direction parallel to the $b$-axis in order to gain the Zeeman energy.
Along the $c$-axis the KS as a function of $H$ starts increasing from the lowest field
and continuously returns to the normal value at around 5T.
Thus the d-vector should be the three components along all  the three directions with complex numbers.
In other words, the SC order parameters must have three dimensional vectorial structure with three components.
This d-vector rotation phenomenon plays a crucial role in understanding the 
field reinforced high field phase as mentioned shortly.

We focus in this paper particularly on the following experiments~\cite{review}:\\
\noindent
(1) The upper critical field $H_{\rm c2}$ is extremely high, reaching $\sim$70T
compared with T$_{\rm c}$=1.6K$\sim$2.0K.\\
\noindent
(2) The $H$-$T$ phase diagram along the magnetic hard $b$-axis consists
of the two phases; low field (LSC) and high field phases (HSC) where in the HSC,
$H_{\rm c2}(T)$ has an unusual positive slope, {\it ie}. $dH_{\rm c2}(T)/dT>0$.\\
\noindent
(3) When tilting $H$ toward the magnetic easy $a$-axis from the $b$-axis
by small angles $\varphi$ up to only $\varphi\sim7^{\circ}$,
the HSC quickly diminishes from the $H$-$T$ phase diagram, leaving the LSC
whose $H_{\rm c2}\sim$10T.\\
\noindent
(4) When the field direction changes from the $b$-axis toward the other magnetic hard $c$-axis
by the angle $\theta$ measured from the $b$-axis, 
the HSC also diminishes up to a little larger angle $\theta\sim12^{\circ}$,
beyond which only the LSC remains.
However, around $\theta\sim35^{\circ}$ the isolated HSC detached from the
LSC appears above the so-called meta-magnetic transition field $H_{\rm m}$
at which the $b$-axis magnetization curve $M_b(H)$ exhibits a jump via a first order phase transition.

Since there is neither quantitative, nor qualitative explanation on those remarkable facts
on UTe$_2$, we try to understand some of these phenomena theoretically in a qualitative level.
In particular, we address the following issues:\\
\noindent
(A) What determines the upper limit of $H_{\rm c2}$?
In a clean limit superconductor~\cite{dirty}, which we assume here,
the orbital limit of $H_{\rm c2}$ without the Pauli paramagnetic effect
is given by $H^{\rm orb}_{\rm c2}=\Phi_0/2\pi \xi^2$ with $\Phi_0$ the flux quantum where
the coherent length $\xi=\hbar v_{\rm F}/\pi T_{\rm c}$.
The Fermi velocity $v_{\rm F}$ measured recently by the dHvA experiment~\cite{aokidHvA}
is $v^{\alpha}_{\rm F}\sim$11.0km/s and $v^{\beta}_{\rm F}\sim$6.3km/s, yielding $H^{\rm orb}_{\rm c2}\sim$12T.
This nicely matches $H_{\rm c2}\sim$10T for the LSC, but is
far less than the observed maximal  $H_{\rm c2}(\theta=35^{\circ})\sim$70T.
Note that according to the $H_{\rm c2}$ analysis by Rosuel et al~\cite{rosuel} and Helm et al~\cite{helm},
the estimated $v_{\rm F}$ in order to explain $H_{\rm c2}\sim$70T is
6.7$\sim$7.1km/s albeit $T_{\rm c}\sim$3K, meaning that the high and low field phases
are governed by the same Fermi surface structure.
Thus we need to understand a mechanism  on what causes the violation of  the orbital depairing limit.\\
\noindent
(B) Why does $H_{\rm c2}$$\parallel$b in the HSC have a positive slope and
terminates abruptly just at $H_{\rm m} $=34T and reappears above $H_{\rm m}$ around $\theta=35^{\circ}$
intermediate between the $b$-axis and $c$-axis~\cite{georg,rosuel,helm}
? Why is it not between the $b$-axis and $a$-axis?

In this paper to address those issues, we assume a spin-triplet pairing 
with a non-unitary form~\cite{ramires} characterized by a complex
d-vector with three components. This non-unitary state quite successfully describes 
not only UTe$_2$, but also
other SC including URhGe and UCoGe. Those are all magnetization-tuned superconductors 
in common~\cite{machida1,machida2,machida3}.

This paper is arranged as follows. First we briefly describe our non-unitary triplet theory 
developed previously~\cite{machida1,machida2,machida3} in the next section II.
In order to understand a mechanism of the violation of the orbital depairing limit of $H_{\rm c2}$
we employ a simple Ginzburg-Landau formalism to illustrate our basic idea as clearly
as possible in Section III. The proposed mechanism is applied to UTe$_2$.
We analyze a variety of experimental data on the $H$-$T$ phase diagrams for various field
orientations in Section IV.  We devote to discussions and perspectives in order to deepen
our understanding on the physics associated with UTe$_2$ and other sister compounds, URhGe and UCoGe.
The topics include the classification scheme of the pairing symmetry, the concept of the d-vector rotation, 
possible chiral-nonchiral transition in high field in  Section V.
Section VI is summary and conclusion.

\section{Theoretical Framework}

\subsection{Preliminaries to Ginzburg-Landau theory}
In order to answer the above questions (A) and (B) 
we start with the most generic Ginzburg-Landau (GL) theory for a spin triplet state. 
Here we briefly summarize our previous theory for further 
developments~\cite{machida1,machida2,machida3}.

We assume 
a non-unitaty A-phase like pairing state described by the complex $\bf d$-vector

\begin{eqnarray}
{\bf d}(k)=\phi(k){\boldsymbol \eta}=\phi(k)({\boldsymbol \eta}'+i{\boldsymbol \eta}'')
\label{d-vector}
\end{eqnarray}

\noindent
(${\boldsymbol\eta}'$ and ${\boldsymbol \eta}''$ are real vectors) among the odd-parity pairing states.
$\phi(k)$ is the orbital part of the pairing function which is not specified in the main part of this paper
because its form is irrelevant for the present arguments. 
The pairing function is classified under the overall symmetry 

\begin{eqnarray}
SO(3)^{\rm spin}\times D_{2h}^{\rm orbital}\times U(1)^{\rm guage}
\label{symmetry}
\end{eqnarray}

\noindent
with the spin, orbital and gauge symmetry respectively.~\cite{machida,annett}.
We assume the weak spin-orbit coupling scheme~\cite{ozaki1,ozaki2}.
This assumption is justified by the fact that 
the d-vector rotation starts from the low fields, $\sim$1T for the c-axis~\cite{ishida3}, and $\sim$5T
for the b-axis~\cite{ishida2}, indicating that the spin-orbit coupling is weak which locks the d-vector 
to crystalline lattices.
This SO(3)$^{\rm spin}$ triple spin symmetry is expressed in terms of a complex
three component vectorial order parameter ${\boldsymbol \eta}=(\eta_a,\eta_b,\eta_c)$.

Under D$_{2h}$$^{\rm orbital}$ symmetry the most general Ginzburg-Landau free energy 
functional up to the quadratic order is expressed
by

\begin{eqnarray}
F^{(2)}=a_0(T-T_{\rm c0}){\boldsymbol \eta}\cdot{\boldsymbol \eta}^{\ast}+
b|{\bf M}\cdot{\boldsymbol \eta}|^2+i\kappa {\bf M}\cdot {\boldsymbol \eta}\times {\boldsymbol \eta}^{\ast}
\label{free2}
\end{eqnarray}

\noindent
with $b$ being a positive constant. 
The last invariant comes from the non-unitarity of the pairing function in the presence of the 
spontaneous moment ${\bf M}(H)$, which is to break the SO(3)$^{\rm spin}$ spin symmetry. 
We assume $\kappa>0$ without loss of generality, but we warn that it could be negative in UTe$_2$.
This term responds to external field directions differently





It is convenient to introduce


\begin{eqnarray}
\eta_{\pm}={1\over \sqrt2}(\eta_b\pm i\eta_c)
\label{eta}
\end{eqnarray}

\noindent
for ${\bf M}=(M_a,0,0)$ where we define the $a$-axis as the magnetic easy axis. 
$\eta_{+}$ ($\eta_{-}$) corresponds to the spin up-up (down-down) pair, or the A$_1$(A$_2$) phase.
Note that the spin quantization axis is defined relative to the ${\bf M}$ direction, namely, the
magnetic easy a-axis here. Due to the magnetic coupling term 
$i\kappa \bf{M}\cdot \boldsymbol \eta\times\boldsymbol \eta^{\ast}$, the spin direction for the
Cooper pair may change.

From Eq.~(\ref{free2}) the quadratic term $F^{(2)}$ becomes

\begin{eqnarray}
F^{(2)}=a_0\{(T-T_{\rm c1})|\eta_{+}|^2+(T-T_{\rm c2})|\eta_{-}|^2\nonumber \\
+(T-T_{\rm c3})|\eta_{a}|^2\}
\label{f2-2}
\end{eqnarray}

\noindent
with 

\begin{eqnarray}
T_{\rm c 1,2}(M_a)=T_{\rm c0} \pm{\kappa\over a_0}M_a, \nonumber \\
T_{\rm c 3}(M_a)=T_{\rm c0} -{b\over a_0}M^2_a.
\label{tc}
\end{eqnarray}

\noindent
Note that the actual second transition temperature is modified to 
$T^{\prime}_{\rm c2}=T_{\rm c0}-({\kappa M_a/ a_0})({{\beta_1-\beta_2})/{2\beta_2}}$
because of the fourth order GL terms~\cite{machida1,machida2,machida3},
but we ignore this correction and maintain the expression of Eq. (\ref{tc}) for clarity of our arguments.

The root mean square average  $\sqrt{\langle M_a^2\rangle}$ of the FM fluctuations along the magnetic easy $a$-axis
is simply denoted by $M_a$ and acts 
to shift the original transition temperature $T_{\rm c0}$ and split it into $T_{c1}$,
$T_{c2}$, and $T_{c3}$ expressed by Eq. (\ref{tc}).
According to this, $T_{c1}$ ($T_{c2}$) increases (decrease) linearly as a function of $M_a$
while $T_{c3}$ decreases quadratically as $M^2_a$ from the degeneracy point $M_a=0$.
The three transition lines meet at $M_a$=0 where the 
three components $\eta_i$ ($i=+,-,a$) are all degenerate. Thus away from the degenerate point at $M_a$=0, the A$_0$
phase starts at $T_{\rm c3}$ quickly disappears from the phase diagram.
Below $T_{\rm c2}$ ($T_{\rm c3}$) the two components $\eta_{+}$ and $\eta_{-}$ coexist, symbolically denoted by
A$_1$+A$_2$. Note that because  their transition temperatures are different,
A$_1$+A$_2$ is not the so-called A-phase which is unitary, but generically non-unitary
except at the degenerate point $M_a$=0 where the totally symmetric phase is realized with 
time reversal symmetry preserved. Thus the A$_1$+A$_2$ phase is the so-called distorted A phase~\cite{3he}. 
Likewise below $T_{\rm c3}$ all the components coexist; A$_1$+A$_2$+A$_0$ realizes.

The magnetic coupling $\kappa$, which is a key parameter to characterize UTe$_2$
in the following, is originally estimated~\cite{mermin} as
$\kappa=T_{\rm c}{N'(0)\over N(0)}ln(1.14\omega/T_{\rm c})$,
with $N'(0)$ the energy derivative of the normal DOS and $\omega$ the energy cut-off.
This term comes from the electron-hole asymmetry near the Fermi level. $\kappa$ indicates
the degree of this asymmetry.
This may be substantial for a narrow band, or the Kondo coherent band in the heavy Fermion
material UTe$_2$. We can estimate
$N'(0)/N(0)\sim 1/E_{\rm F}$ with the Fermi energy $E_{\rm F}$.
Because $T_{\rm c}$=2mK and $E_{\rm F}$=1K in $^3$He, $\kappa\sim10^{-3}$,
while for UTe$_2$ $T_{\rm c}\sim$1K and $E_{\rm F}\sim T_{\rm K}$ with the 
Kondo temperature $T_{\rm K}\sim$30K~\cite{review}.
$\kappa\sim10^{-1}$. 
We also note that the sign of $\kappa$ can be either positive or negative,
depending on the detailed energy dependence at the Fermi level because it is $\propto N'(0)$.
If $\kappa>0$ ($\kappa<0$), the up-up (down-down) pair appears at higher $T$.
Thus the Knight shift remains unchanged (decreases) below $T_{\rm c1}$.

In the following discussions we consider the case where the two components $\eta_{+}$ and $\eta_{-}$
are nonvanishing, ignoring the third component $\eta_{a}$ since 
under ambient pressure UTe$_2$ exhibits the two phases LSC and HSC, corresponding to 
 $\eta_{+}$ and $\eta_{-}$ respectively.
 Note, however, that under pressure the third component becomes relevant~\cite{machida2}.
 We redefine the notation $\kappa/a_0\rightarrow \kappa$ from now on.
 
\section{Upper critical field}


Under an applied field with the vector potential $\bf A$, the gradient GL energy is given 
under D$_{2h}$$^{\rm orbital}$ symmetry


\begin{eqnarray}
F_{grad}=\sum_{\nu=a,b,c}\{K_a|D_x\eta_{\nu}|^2+K_b|D_y\eta_{\nu}|^2+K_c|D_z\eta_{\nu}|^2\}
\label{gradient}
\end{eqnarray}

\noindent
where $K_{a}$, $K_{b}$, and $K_c$ are the effective mass along the $a$ $b$, and $c$-axes.
$D_i=-i\nabla_i+{2\pi\over \Phi_0}A_i$ is the gauge invariant derivative with
$\Phi_0$ being the quantum flux and $A_i$ the vector potential component.
We emphasize as seen from this form of Eq.~(\ref{gradient}) 
that $H_{\rm c2}$ for the three components each starting at $T_{\rm c j}$ ($j=1,2, 3$) intersects
each other, never avoiding or leading to a level repulsion. The level repulsion may occur 
for the pairing states belonging to multi-dimensional representations
 (see for example [\onlinecite{repulsion1,repulsion2,repulsion3,repulsion4}] in UPt$_3$).
The external field $H$ comes in also through $M_a(H)$ in addition to the vector potential $\bf A$
which gives rise to the orbital depairing.

Thus each component is independent within the quadratic terms.
The Ginzburg-Landau free energy density $F$ under external magnetic field $H$
in terms of the SC order parameter $\eta_{\pm}$ given by

\begin{eqnarray}
F&=&\sum_{i=\pm}\{a_0(T-T_{{\rm c},i})|\eta_{i}|^2  \nonumber \\ 
& &+K_{a}|D_x\eta_i|^2+K_{b}|D_y\eta_i|^2+K_c|D_z\eta_i|^2\}.
\label{GL}
\end{eqnarray}

\noindent
The variation with respect of $\eta_i^{\ast}$ leads to the Ginzburg-Landau equation

\begin{eqnarray}
a_0(T-T_{\rm c})\eta_i+(K_{a}D_x^2+K_{b}D_y^2+K_cD_z^2)\eta_i=0.
\label{osci}
\end{eqnarray}

\noindent
Following the standard procedure~\cite{tinkham},
the upper critical field $H_{\rm c2}$ is obtained as the lowest eigenvalue of the linearized 
Ginzburg-Landau equation, or Schr\"odinger type equation of a harmonic oscillator, namely,


\begin{eqnarray}
H^{(+)}_{{\rm c2},j}(T)=\alpha^j_0(T_{\rm c0} +\kappa M_a-T)\nonumber \\ 
H^{(-)}_{{\rm c2},j}(T)=\alpha^j_0(T_{\rm c0} -\kappa M_a-T)
\label{hc2}
\end{eqnarray}

\noindent
with $j$=$a$, $b$ and $c$-axis. We have introduced the coefficients,

\begin{eqnarray}
\alpha^{a}_0&=&{\Phi_0\over 2\pi \sqrt{K_bK_{c}}}a_0,\qquad
\alpha^{b}_0={\Phi_0\over 2\pi \sqrt{K_cK_{a}}}a_0,\nonumber \\ 
\alpha^{c}_0&=&{\Phi_0\over 2\pi \sqrt{K_aK_{b}}}a_0.
\label{mass}
\end{eqnarray}

\noindent
Those coefficients determine the initial slopes of the upper critical fields.
$H^{(+)}_{{\rm c2},j}$ and $H^{(-)}_{{\rm c2},j}$ are the upper critical fields for 
the spin up-up and down-down pair, or the A$_1$ and A$_2$ phase respectively.

The above equation (\ref{hc2}) is cast into a generic form:

\begin{eqnarray}
H_{{\rm c2}}-\alpha_0\kappa M(H_{\rm c2})=\alpha_0(T_{\rm c0}-T).
\label{gl}
\end{eqnarray}

\noindent
The right hand side of Eq.(\ref{gl}) is nothing but 

\begin{eqnarray}
H_{\rm c2}^{\rm orb}(T)=\alpha_0(T_{\rm c0}-T)
\label{heff}
\end{eqnarray}

\noindent
for unperturbed upper critical field due to the orbital depairing limit
with $T_{\rm c0}$ whose maximum value is given by $H_{\rm c2}^{\rm orb}(T=0)=\alpha_0T_{\rm c0}$.
On the left hand side of Eq.(\ref{gl}) we define the effective field

\begin{eqnarray}
H_{\rm eff}=H_{{\rm ext}}-\alpha_0\kappa M(H_{\rm ext}).
\label{heff}
\end{eqnarray}

\noindent
This implies that the external field $H_{{\rm ext}}$ is reduced by the amount of 
$\alpha_0\kappa M(H_{\rm ext})$. 
The upper bound of the orbital depairing field of $H_{\rm c2}^{\rm orb}(T)$
for the $a$-axis, for example, is determined by 

\begin{eqnarray}
H_{\rm c2}^{\rm orb}(T\rightarrow0)=
\alpha^a_0T_{\rm c0}={\Phi_0\over 2\pi \sqrt{K_bK_{c}}}a_0T_{\rm c0}.
\label{h0}
\end{eqnarray}

\noindent
This is given in turn by the expression  in the clean limit: $H_{\rm c2}^{\rm orb}(T)={\Phi_0/ 2\pi \xi^2}$ 
with the coherence length $\xi={\hbar v_{\rm F}/ \pi T_{\rm c0}}$.
Namely, at $H_{\rm c2}^{\rm orb}(0)$ the inter-vortex distance becomes comparable to
the core size $\xi$. This gives rise the absolute value of the upper limit of $H_{\rm c2}^{\rm orb}(0)$
in general.
In order to break this absolute upper limit due to the orbital depairing,
 the effective magnetic field $H_{\rm eff}$
must be reduced from the external field $H_{{\rm ext}}$.
This idea is the same as in the case developed for a spin singlet pairing~\cite{ce}
and somewhat similar to the so-called Jaccarino-Peter mechanism~\cite{jp}.
From now on we surpress subscript ``ext'',
thus $H_{\rm ext}\rightarrow H$.

It is clear to see that at $T=0$ the absolute value of $H_{\rm eff}$ is bounded by

\begin{eqnarray}
|H_{\rm c2}-\alpha_0\kappa M(H_{\rm c2})|\leq H^{\rm orb}_{\rm c2}(T=0)=\alpha_0T_{\rm c0}
\label{bound}
\end{eqnarray}

\noindent
for $H_{\rm c2}(0)$ to be a solution.
Thus  $H_{\rm c2}(0)$ could be enhanced at $T\rightarrow 0$.

\begin{figure}
\includegraphics[width=8cm]{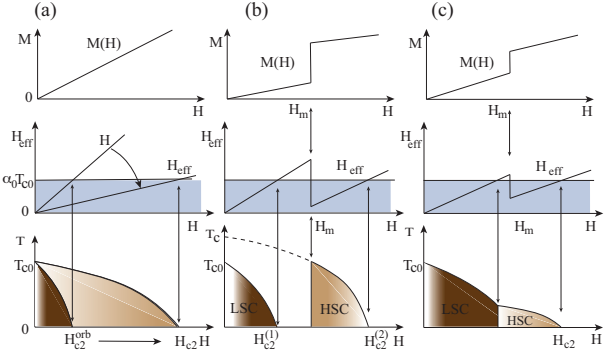}
\caption{(a) In the case of magnetization curve $M(H)=\chi H$
(upper panel). $H_{\rm eff}$ is reduced compared with the external field.
The allowed region with grey color bounded by $\alpha_0T_{\rm c0}$ extends to a higher field (middle panel).
$H_{\rm c2}$ is enhanced compared with $H^{\rm orb}_{\rm c2}$ (the bottom panel).
(b) When the magnetization has the jump at the metamagnetic field $H_{\rm m}$,
$H_{\rm eff}$ becomes outside of the allowed region at  $H^{(1)}_{\rm c2}$.
But it comes back above $H_{\rm m}$ and HSC appears, separated from LSC. 
The extrapolated $T_{\rm c}$ for HSC is higher than $T_{\rm c0}$ for LSC (dotted curve in the bottom panel).
(c) The metamagnetic jump is smaller than the case (b).
LSC and HSC are overlapped to appear. The grey regions in the middle panels in (a), (b), and (c) 
show the allowed region for $H_{\rm c2}$.
}
\label{f0}
\end{figure}

Let us now examine the typical cases for several magnetization curves
as shown in Fig.~\ref{f0}.
We first consider the simplest case where
the magnetization curve is given by $M(H)=\chi H$ as displayed in the upper panel of Fig.~\ref{f0}(a).
Since $H_{\rm eff}$ is reduced by the presence of $M(H)$ in Eq.~(\ref{heff}) (the middle panel in Fig.~\ref{f0}(a)),
we find

\begin{eqnarray}
H_{\rm c2}(T)={H^{\rm orb}_{\rm c2}(T)\over 1-\alpha_0\kappa\chi}
\label{henhance}
\end{eqnarray}

\noindent
$1-\alpha_0\kappa\chi$ is the enhancement factor relative to $H^{\rm orb}_{\rm c2}(T)$
(the bottom panel in Fig.~\ref{f0}(a)).
Thus in principle $H_{\rm c2}(T)$ increases indefinitely toward the critical point
$\alpha_0\kappa\chi=1$ from below.
As a general tendency, when the magnetization becomes saturated at higher field,
 $H_{\rm c2}(T)$ eventually tends to be finite.
 
 Next we consider  the case where 
 the magnetization curve has a jump at the metamagnetic transition at $H_{\rm m}$
 as shown in the upper panel of Fig.~\ref{f0}(b).
$H_{\rm eff}$ exceeds the allowed maximum region set by 
 $\alpha_0T_{\rm c0}$ in Eq.~(\ref{heff}) at a lower field
 as shown in the middle panel of Fig.~\ref{f0}(b), thus low SC (LSC) phase is terminated
at  $H^{(1)}_{\rm c2}(T)$ (see the bottom panel of Fig.~\ref{f0}(b)). 
However, just above $H_{\rm m}$, $H_{\rm eff}$ enters again the
allowed region with grey color in the middle panel, thus high SC (HSC) 
appears from $H_{\rm m}$ to $H^{(2)}_{\rm c2}(T)$ as shown in the bottom panel of Fig.~\ref{f0}(b).
In this case LSC and HSC are separated in $H$-$T$ phase diagram shown in the bottom panel in
Fig.~\ref{f0}(b). 

 Depending on the magnetization curve with the metamagnetic transition,
 the different situation may occur as shown in Fig.~\ref{f0}(c). 
 Since $H_{\rm eff}$ defined in Eq.~(\ref{heff}) 
 is determined by the combination of $M(H)$ and the coupling constant $\alpha_0\kappa$, 
two SC phases of LSC and HSC are overlapped as shown in the bottom panel of Fig.~\ref{f0}(c).
 This is contrasted with the case mentioned above where LSC and HSC are separated by the 
 normal state along the $H$ axis in $H$-$T$ phase diagram.
 Notice that in those examples LSC and HSC are the same pairing state.

\section{Analysis of $H_{\rm c2}(T)$ in $\rm{UTe}_2$}
\subsection{H//b}

\begin{figure}
\includegraphics[width=6cm]{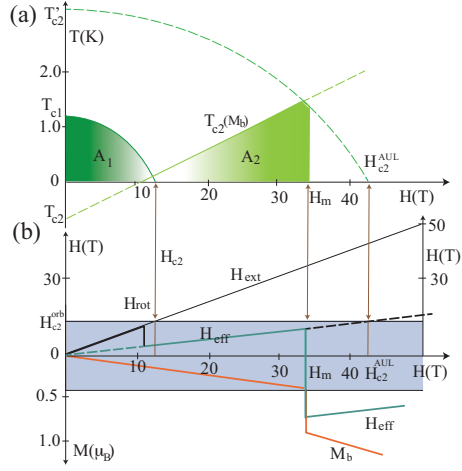}
\caption{(a) The resulting $H$-$T$ phase diagram with the A$_1$ (LSC) and A$_2$ (HSC) phases.
The dashed lines are not realized.
(b) The constructed $H_{\rm eff}$ (green curve) at $T=0$ as a function of the
external field $H$ using the measured magnetization curve~\cite{miyake} 
of $M_b(\mu_{\rm B})$ (red curve). $H_{\rm rot}$ is the d-vector rotation field.
$H_{\rm m}$ is the metamagnetic transition field. $H^{\rm AUL}_{\rm c2}$
is the absolute upper limit of $H_{\rm c2}$.
}
\label{f1}
\end{figure}

In this section, we examine the $H$-$T$ phase diagram in UTe$_2$ for $H\parallel b$
by applying the previous general considerations based on GL theory for non-unitary pairing.
In order to explain various mysteries associated with the phase diagram for $H\parallel b$,
it is essential to know the magnetization curve $M(H)$ in $H\parallel b$.
According to the measurement by Miyake et al~\cite{miyake}
$M(H)$ has the metamagnetic transition at $H_{\rm m}$=34T via a first order
with the large magnetization jump, which is shown by the red curve of $M_b$ in Fig.~\ref{f1}(b).
Accordingly, the effective field $H_{\rm eff}$ shown by the green curve there
exhibits a sharp drop at $H_{\rm m}$. By choosing an appropriate parameter value 
for $\alpha_0\kappa$ which is only the adjustable parameter in our theory,
we can reproduce the experimental data.
Namely, $H_{\rm eff}$ is reduced below $H<H_{\rm m}$ as seen by the green curve of Fig.~\ref{f1}(b).
However, beyond $H_{\rm m}$ it exceeds the limit of the allowed region denoted by the grey band.

In Fig.~\ref{f1}(a) the A$_1$ phase starts at $T_{\rm c1}$ and disappears at a lower field
because the Cooper pair polarization points to the $a$-axis evidenced by the KS
experiment~\cite{ishida1,ishida2,ishida3,ishida4,ishida5}. In low fields KS remains unchanged (drops) for the
$a$-axis ($b$- and $c$-axis) field. Thus for the A$_1$ phase $H_{\rm eff}=H$ 
because of ${\bf d}\times{\bf d}^{\ast} \perp M_b$.

On the other hand, the A$_2$ phase with the increasing $T_{\rm c2}$
changes the d-vector direction during the d-vector rotation for the field range 5T$\sim$12T
in order that ${\bf d}\times{\bf d}^{\ast} \parallel M_b$, thus now
$T_{\rm c2}=T_{\rm c0}+\kappa M_b(H)$ instead of $M_a$ originally 
given in Eq.~(\ref{tc}), or 
$T_{\rm c2}$ increases with $M_b(H)$ as shown in Fig.~\ref{f1}(a).
However, even if $T_{\rm c2}$ is increasing indefinitely, the A$_2$ phase ceases to exist above 
$H_{\rm m}$ because $H_{\rm eff}$ exceeds the limit. It terminates at $H^{\rm AUL}_{\rm c2}$
where as shown in the dotted line of Fig.~\ref{f1}(b) the extrapolated $H_{\rm eff}$ from below
exceeds the limit. This defines the absolute upper limit of   $H_{\rm c2}$, or  $H^{\rm AUL}_{\rm c2}$,
which is given by
$H^{\rm AUL}_{\rm c2}=\alpha_0(T'_{\rm c2}-T)$ where $T'_{\rm c2}$ is not realized.
Note that  as seen from Fig.~\ref{f1}(a) a part of $H^{\rm AUL}_{\rm c2}(T)$ is realized
where $H_{\rm eff}$ is still within the allowed region.
Those constitute the whole A$_2$ phase shown in Fig.~\ref{f1}(a).

\subsection{$b$ to $a$}

\begin{figure}
\includegraphics[width=7cm]{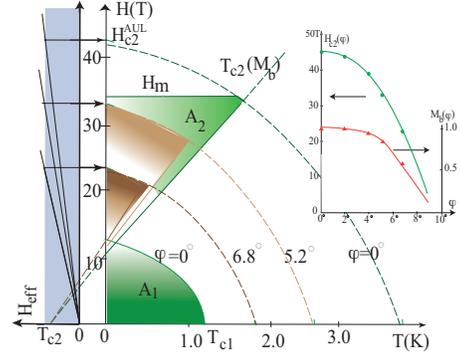}
\caption{The phase diagram in $H$-$T$ plane for $H\parallel b$ and the field orientations tilted 
by the angle $\varphi$ measured from the $b$-axis toward the $a$-axis. 
The A$_2$ or HSC quickly shrinks as $\varphi$
increases while A$_1$ or LSC remains almost unaffected.
$H^{\rm AUL}_{\rm c2}(\varphi)$ becomes low as $\varphi$ increases indicated by left hand side
because the projection of $M_b(\varphi)$ strongly decreases as postulated in the inset.
The resulting upper critical field $H_{\rm c2}(\varphi)$ is shown there.
}
\label{f2}
\end{figure}

When the magnetic field is tilted from the magnetic hard $b$-axis toward the
magnetic easy  $a$-axis by the angle $\varphi$ measured from $b$,  the
HSC phase quickly diminishes from the $H$-$T$ phase diagram up to $\varphi\sim7^{\circ}$
while LSC remains the same.
In order to understand this intriguing behaviors, we apply the same idea above
by postulating the $M_b(\varphi)$ as a function of  $\varphi$.
When tilting the field direction away from the $b$-axis, $M_b(\varphi)$ 
generally decreases because $M_b$ component projecting onto the
field direction becomes small.
Therefore, $H_{\rm eff}(\varphi)$ increases with $\varphi$
as shown in the left hand side of Fig.~\ref{f2}, implying that 
$H^{\rm AUL}_{\rm c2}(\varphi)$ is lowered. The resulting $H^{\rm AUL}_{\rm c2}(\varphi)$
is plotted by the dotted curves in Fig.~\ref{f2} for the selected angles.
Since $T_{\rm c2}(\varphi)=T_{\rm c0}+\kappa M_b(\varphi)$ becomes sharper to rise
or $T_{\rm c2}(\varphi)$ at $T_{\rm c2}$ rotates counterclockwise 
as depicted in Fig.~\ref{f2}, the A$_2$ regions with the triangle areas (brown color)
become shrink and disappears from the $H$-$T$ phase diagram.

The postulated $M_b(\varphi)$ behavior in order to reproduce the phase diagram
is depicted in the inset of Fig.~\ref{f2}, which is far from that expected by simple projection
of $M_b(\varphi)$ onto the field direction.
$M_b(\varphi)$ decreases quickly upon tilting by a few degrees, which is quite noteworthy.
This might be understandable because the magnetic easy $a$-axis is special;
The moment $M_b$ tends to redirect toward the easy $a$-axis in order to gain the
magnetic energy by increasing the $M_a$ component, thus the rotation of the moment direction 
of $M_b$ may be larger than the simple projection count.
A similar large change of the magnetization curve by small tiltings of the field direction
 is observed in URhGe from the hard to easy axis case~\cite{nakamura}.
Reflecting the strong decrease of $M_b(\varphi)$, the resulting $H_{\rm c2}(\varphi)$
sharply drops as depicted in the inset of Fig.~\ref{f2}.

\subsection{$b$ to $c$}

We examine the phase diagram for the field orientation tilted from the
$b$-axis to the other hard axis $c$-axis by the angle $\theta$ measured 
from the $b$-axis to understand the isolated HSC phase whose 
maximum $H_{\rm c2}$ reaches $\sim$70T far above the orbital depairing 
upper critical field.

\begin{figure}
\includegraphics[width=7cm]{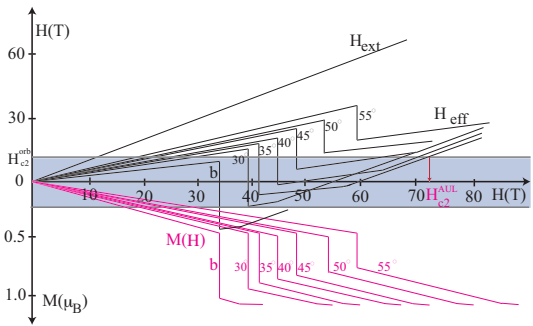}
\caption{The magnetization curve $M_b(H)$ (the red curve) for $H\parallel b$-axis obtained experimentaly~\cite{miyake}. 
The other magnetization curves for various angles of $\theta$ are reconstructed
by projecting $M_b(H)$ onto the magnetic field direction.
$H_{\rm eff}(\theta)=H-\alpha_0\kappa M(\theta)$ is constructed from $M(\theta)$ thus obtained.
The grey color band at the center indicates the allowed region for $H_{\rm c2}$.
The intersection point between $H_{\rm eff}(\theta)$ and the grey band yields the
absolute upper limit $H^{\rm AUL}_{\rm c2}$.
}
\label{f3}
\end{figure}

Let us start to evaluate the magnetization curves $M_b(\theta)$ for the arbitrary angle $\theta$,
which is a key quantity to determine $H_{\rm c2}$.
It is rather easy to reconstruct $M_b(\theta)$ from the magnetization curve $M_b$
which is measured~\cite{miyake} since we know the experimental fact that $H_{\rm m}\propto 1/\cos\theta$.
This means that the projection of $M_b$ onto the field direction determines the
magnetization curve for $M_b(\theta)$.
Therefore, by projecting $M_b$ onto the field direction we obtain $M_b(\theta)$ 
for arbitrary angle. In Fig.~\ref{f3}, $M_b(\theta)$ is depicted as the red curves for
the relevant angles of $\theta$.
We can check this procedure for the experimental data $M_b(\theta\sim23^{\circ})$ for 
$H\parallel (011)$-direction~\cite{miyake} by subtracting the contribution from 
the magnetization component along the $c$-axis $M_c(\theta)$.

Using those magnetization curves and the same parameter value for $\alpha_0\kappa$,
we obtain  $H_{\rm eff}(\theta)=H-\alpha_0\kappa M(\theta)$ as shown in Fig.~\ref{f3}.
It is seen from this that for $\theta\geq 30^{\circ}$ the lower edge of $H_{\rm eff}(\theta)$
begins entering the allowed region, yielding the HSC up to $H^{\rm AUL}_{\rm c2}$.
Upon further increasing $\theta$, $H_{\rm eff}(\theta)$ is leaving this region, thus
there is no HSC for $\theta>50^{\circ}$.

\begin{figure}
\includegraphics[width=6cm]{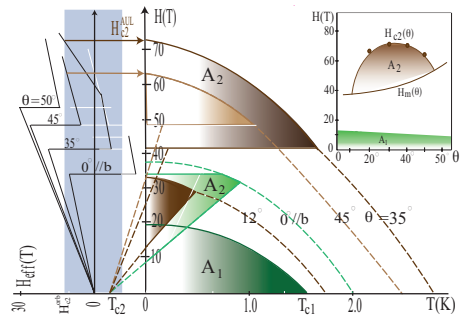}
\caption{The $H$-$T$ phase diagram for various $\theta$, including the case $H\parallel b$-axis
for comparison. The allowed region of $H_{\rm eff}$ is displayed in the left hand side by grey color,
which is the same as in Fig.~\ref{f3}.
For $\theta=35^{\circ}$ the HSC (A$_2$) is allowed for $H_{\rm m}<H_{\rm eff}<H^{\rm AUL}_{\rm c2}$.
The $H_{\rm c2}$ curve starts at $H^{\rm AUL}_{\rm c2}$ toward $T_{\rm c2}$ at $H=0$.
But the HSC terminates abruptly at $H_{\rm m}$ below which $H_{\rm eff}$ is outside of
the allowed region. The allowed region at the low field is not available for the A$_2$
because $T_{\rm c2}<0$ there. It is used by the LSC (A$_1$), which is relatively
unchanged in varying $\theta$, including the case $H\parallel b$-axis.
The inset shows the HSC (A$_2$) and LSC (A$_1$) as a function of $\theta$.
}
\label{f4}
\end{figure}

We can construct the $H$-$T$ phase diagram for $\theta$ shown in Fig.~\ref{f4}
where the selected $\theta$ cases are displayed, including $H\parallel b$-axis
for comparison.
The left side bar denotes $H_{\rm eff}(\theta)$ explained above.
For $\theta=12^{\circ}$ the A$_2$ phase barely remains beyond which
there is no trace of the A$_2$ phase  below $H_{\rm m}$ in the phase diagram.
This is because $T_{\rm c2}(M_b)$ curves (denoted in the dotted straight lines in Fig.~\ref{f4})
starting at $T_{\rm c2}$ for $H=0$ rotate counter-clockwise due to the decrease of 
the $M_b$ projection.
However for $\theta=35^{\circ}$ this  $T_{\rm c2}$ line still reaches the metamagnetic transition
field, which allows the HSC to exist above $H_{\rm m}$ as shown in Fig.~\ref{f4}.
Thus starting from $H^{\rm AUL}_{\rm c2}\sim70$T through $H_{\rm m}$, the $H_{\rm c2}$ curve is extended
toward $T_{\rm c2}\sim$3K at $H$=0. However the actual HSC phase disappears abruptly
at $H_{\rm m}$ because $H_{\rm eff}$ is outside of the allowed region below $H_{\rm m}$.
For $\theta=45^{\circ}$ since there is only tiny field region allowed for $H_{\rm eff}$
as seen from the left hand side of Fig.~\ref{f4}, the resulting HSC region in the $H$-$T$ phase diagram
shrinks. There is no HSC allowed for $\theta=50^{\circ}$.
Those features are displayed in the inset of Fig.~\ref{f4}.

\section{Discussions ans perspectives}

\subsection{Parameter value of $\alpha_0\kappa$}

We examine the parameter values used in this paper.
The key parameter in this work is the product $\alpha_0\kappa$
of $\alpha_0$ introduced in Eq.~(\ref{mass}) and $\kappa$
defined in Eq.~(\ref{free2}). We ignore the small anisotropy of the 
initial slopes of  $H_{\rm c2}$ at $T_{\rm c}$
for three field orientations of the $a$-, $b$-, and $c$-axes.
From the initial slopes, we find $\alpha_0$=12T/1.6K=7.5T/K.
From the previous estimate $\kappa$=6.9K/$\mu_{\rm B}$~\cite{machida3}, which is determined
by the splitting between $T_{\rm c1}$ and $T_{\rm c2}$
and the amplitude of the ferromagnetic fluctuation moment
along the $a$-direction.
We obtain $\alpha_0\kappa$=51.8T/$\mu_{\rm B}$.
From Eq.~(\ref{heff}) it is seen that

\begin{eqnarray}
H_{\rm eff}=H-\alpha_0\kappa M=H-J_{\rm cf}M,
\label{heff2}
\end{eqnarray}

\noindent
namely,  this combination is nothing but the form of the exchange integral $J_{\rm cf}$
between the 5f localized moment and conduction electrons, {\it ie}. $J_{\rm cf}=\alpha_0\kappa$.

It is interesting to notice the case in the recently found heavy Fermion superconductor 
CeRh$_2$As$_2$~\cite{khim} where $H^c_{\rm c2}$=16T and $T_{\rm c}$=0.35K.
This compound is known to break the Pauli-Clogston limit $H_{\rm p}$=1.84$T_{\rm c}\sim$0.6T by far.
 In order to overcome this Pauli-Clogston limit  for this spin singlet superconductor,
 we introduce the effective field 
 $H_{\rm eff}=H-J_{\rm cf}M$ 
 where the internal field is exerted from the localized 4f moment $M$ to cancel the external applied field~\cite{ce}.
 The exchange integral is estimated as  
$J^c_{\rm cf}$=52.5T/$\mu_{\rm B}$ ($J^{ab}_{\rm cf}$=23.4/$\mu_{\rm B}$) for the $c$-axis  
($ab$-plane) in tetragonal crystal. Those numbers remarkably coincide with the present system, but it may
be only coincident.
The important thing is that to achieve the high $H_{\rm c2}$
it is necessary to break the Pauli-Clogston limit for a spin singlet superconductor
or the orbital depairing limit for a spin triplet superconductor.
Here we propose a common mechanism where the external field is effectively
cancelled by the internal field due to the moments of the localized f electrons through the
exchange coupling to the itinerant  electron system.

\subsection{Pairing symmetry of UTe$_2$ and classification scheme}

The present analysis clearly shows that the non-unitary state in the chiral form
$d(k)=(b+ic)(k_b+ik_c)$ is best suitable for UTe$_2$. Here we chose the orbital part $\phi(k)=k_b+ik_c$.
Under applied fields the d-vector rotates
so as to save the Zeeman energy. This means that the spin-orbit coupling to lock the
d-vector to the underlying crystal lattices
 is weak and finite. Namely, the d-vector rotation fields $H_{\rm rot}$ depend on the field
orientation, that is, $H_{\rm rot}=5\sim12$T for 
$H\parallel b$-axis and $H_{\rm rot}=1$T for 
$H\parallel c$-axis. Those weak fields of $H_{\rm rot}$ indicate the strength of the spin-orbit coupling (SOC).
Therefore we have to resort to the weak SOC scheme for the pairing symmetry classification.

The spin-orbit coupling is anisotropic, thus the spin space symmetry for the
Cooper pairs is weakly broken from the original SO(3)$^{\rm spin}$. Furthermore,
the slow ferromagnetic fluctuations also break it to split the SC transition temperature
into three, $T_{\rm c1}$, $T_{\rm c2}$, and $T_{\rm c3}$. In this way we can reasonably identify
the relevant Cooper pair symmetry started from SO(3)$^{\rm spin}$, which is decoupled
with the orbital part of the pairing function in this scheme. We emphasize that since
in the strong SOC case advocated by others~\cite{anderson,blount,gorkov,taillefer} 
the spin space symmetry and the orbital space symmetry are tightly coupled,
there is no freedom to allow the d-vector rotation. 
As mentioned above the gradual rotation of the d-vector via a second order phase transition
is accounted for only by the weak SC case.
As for the orbital symmetry governed by the crystalline symmetry D$_{\rm 2h}$,
there is no multi-dimensional representation. Thus the choice of the chiral form
$k_b+ik_c$ which is consistent with many experiments~\cite{review,metz,kittaka} is ad hoc at this stage.
It may be that the classification scheme based on the D$_{\rm 2h}$ crystalline symmetry turns out to
be irrelevant and more larger symmetry group is needed.
Note that a convex curve behavior of the Sommerfeld coefficient $\gamma(H)$ 
at low fields for $H\parallel b$-axis associated with the Pauli paramagnetic effect~\cite{pauli} 
is an important signature of the d-vector locking and should be checked experimentally.

\subsection{d-vector rotation}
The d-vector rotation is an important concept for describing the phenomena
associated with peculiar $H$-$T$ phase diagrams. In particular
for $H\parallel b$-axis the positive slope above $H\simeq$12T can be accounted for
by the d-vector rotation where the d-vector becomes perpendicular to the $b$-axis
so that the magnetic coupling $i{\bf M}_b\cdot{\bf d}\times\bf{d^{\ast}}$ is active and 
fully takes advantage from this magnetic energy,
otherwise this invariant does not help to raise $T_{\rm c2}$.
In this sense the d-vector rotation is essential to capture this phenomenon.

Microscopically the d-vector rotation occurs as a change of the spin texture
formed by the spatial modulation of the three dimensional d-vector, or
the Cooper pair spin polarization defined by ${\bf S}(r)=i{\bf d}\times\bf{d^{\ast}}$.
The averaged ${\bf S}(r)$ over the vortex unit cell determines the direction of the
d-vector. The d-vector rotation is induced because the competition between the Zeeman energy and
the pinning of the d-vector to the underlying lattices due to the SOC.
A microscopic theory based on quasi-classical Eilenberger equation
is now in progress where intriguing spin textures, including a pair of the half-quantized vortices
and Majorana zero modes both with spinless and spinfull are stabilized~\cite{tsutsumi}.

\subsection{Chiral-nonchiral transition and $\beta$ phase}

When the magnetic field $H\parallel b$-axis is applied to the fully polarized
nonunitary chiral $p$ state $(a+ic)(k_b+ik_c)$,
the chiral-nonchiral transition may occur. This mechanism is originally proposed
by Scharnberg-Klemm~\cite{klemm}.
This is simply because to compare the two upper critical fields for the chiral state $(a+ic)(k_b+ik_c)$
and the nonchiral state $(a+ic)k_b$ the latter has higher $H_{\rm c2}$ in general,
a factor $\sim$1.5 higher for the spherical Fermi surface~\cite{miranovic}.  
The line node in $(a+ic)k_b$ is robust under fields compared with $(a+ic)(k_b+ik_c)$ having the
point nodes.
This nonchiral state $(a+ic)k_b$ is named as the so-called $\beta$ phase~\cite{3he,ozaki1,ozaki2}.
The $\beta$ phase produced by high magnetic fields from the polar phase is
 recently identified in superfluid $^3$He confined in nematic aerogel~\cite{beta}.
Thus it is quite interesting to investigate this possibility further in our superconductor.
We have already identified the A$_1$, A$_2$, A$_1$+A$_2$ (distorted A), and
A$_1$+A$_2$+A$_0$ phases in lower and intermediate field regions 
under ambient pressure and under pressure respectively~\cite{machida1,machida2,machida3}.

\subsection{Application to URhGe and UCoGe}

\begin{figure}
\includegraphics[width=6cm]{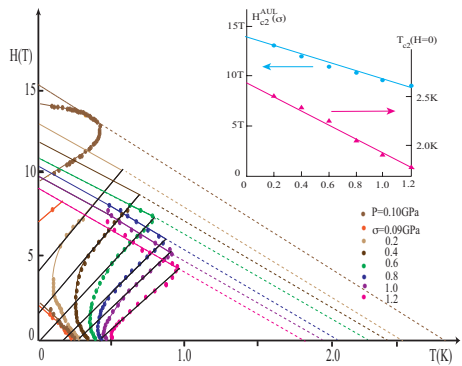}
\caption{$H$-$T$ phase diagram under hydrodynamic (P) and uniaxial ($\sigma$) pressure
for URhGe. The extrapolated straight line  to lower $T$ defines $H^{\rm AUL}_{\rm c2}$ and
$T_{\rm c2}(H=0)$ to higher $T$  respectively. The pressure dependences of $H^{\rm AUL}_{\rm c2}$ and
$T_{\rm c2}(H=0)$ are shown in the inset, indicating the linear scaling for both quantities with the
linear decrease of $M_c$. The dotted points are the experimental data~\cite{urhge1,urhge2}.
}
\label{f5}
\end{figure}

In order to examine the validity of the present theory, we apply it to
other materials, ferromagnetic superconductors URhGe and UCoGe
which are best systems to check our idea.
Under hydrodynamic and uniaxial pressure the $H$-$T$ phase diagrams  in URhGe
continuously change as shown in Fig.~\ref{f5}.
The features are strikingly similar to those we have just seen, 
such as 

\noindent
(1) $T_{\rm c}(H)$ increases as $H$ increases in some part of $H$-$T$ phase diagram,

\noindent
(2) the extrapolated $T_{\rm c}$ from the high field $H_{\rm c2}$ to
high $T$ exceeds $T_{\rm c0}$ at $H=0$, 

\noindent
(3) the HSC is separated from
LSC at low pressure,

\noindent
(4) HSC and LSC is overlapped in high
pressure regon.

Let us examine those features observed in URhGe in light of the present idea.
It is known that under uniaxial pressure $\sigma$ the spontaneous moment $M_c$
decreases linearly and vanishes at $\sigma$=1.2GPa, namely
$M_c(\mu_{\rm B})=0.4-0.33\sigma$(GPa).
It is reasonable to consider that $M_b(H)=\chi_b H$
where $\chi_b$ decreases in proportion with $\sigma$, namely
$\chi_b=\chi_{b0}-A\sigma$ with $A$ positive constant because the spontaneous moment
$M_c$ sets the overall magnetic scale.
Thus it is expected that $H^{\rm AUL}_{\rm c2}$  is given
by $H_{\rm c2}-\alpha_0\kappa M_c=\alpha_0T_{\rm c0}$, or

\begin{eqnarray}
H_{\rm c2}-\alpha_0\kappa \chi_bH_{\rm c2}=\alpha_0T_{\rm c0}, \nonumber\\
H_{\rm c2}-\alpha_0\kappa (\chi_{b0}-A\sigma )H_{\rm c2}=\alpha_0T_{\rm c0}.
\label{sigma}
\end{eqnarray}

\noindent
The above Eq.~(\ref{sigma}) is rewritten as

\begin{eqnarray}
{H_{\rm c2}(\sigma)\over H_{\rm c2}(\sigma=0)}={1\over{1+{\kappa\alpha_0A\sigma\over {1-\kappa\alpha_0\chi_0}}}}\simeq1-{\kappa H_{\rm c2}(\sigma=0)A\sigma\over T_{\rm c0}}.
\label{sigma2}
\end{eqnarray}

\noindent
Namely, $H^{\rm AUL}_{\rm c2}(\sigma)$ decreases linearly with $\sigma$.
This also implies that $T_{\rm c2}(H=0)$ decreases linearly with $\sigma$.
As displayed in the inset of Fig.~\ref{f5} this relation is well obeyed.

\begin{figure}
\includegraphics[width=5cm]{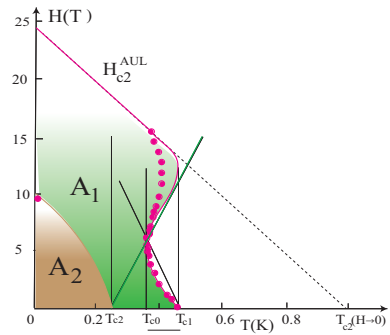}
\caption{$H$-$T$ phase diagram~\cite{machida3} 
for $H\parallel$ $b$-axis in UCoGe. 
The extrapolated straight line  to lower $T$  and higher $T$ defines $H^{\rm AUL}_{\rm c2}=$24T and
$T_{\rm c2}(H=0)=1.0$K respectively. 
The red dots are the experimental data~\cite{ucoge1,ucoge2}.
}
\label{f6}
\end{figure}

Here we quote our previous figure~\cite{machida3} on UCoGe modified slightly as Fig.~\ref{f6}.
It is clear that there certainly exists $H^{\rm AUL}_{\rm c2}$ in UCoGe too.
The extrapolated $T_{\rm c2}(H\rightarrow0)\sim$1.0K far higher than $T_{\rm c1}$.
The S-shaped $H_{\rm c2}$ is limited from the above, evidencing the presence of $H^{\rm AUL}_{\rm c2}$.
We now understand the reason why it is so.

\subsection{Perspectives}

The present material UT$_2$ is considered to be nearly ferromagnetic
although the ``static'' long range ferromagnetic (FM) ordering is absent~\cite{sonier}.
The slow FM fluctuations are reported by several experiments~\cite{tokunaga1,tokunaga2,sonier,furukawa}.
This situation is similar to UPt$_3$ where the antiferromagnetic (AF) order above $T_{\rm c}$ is not truly
static and long-ranged order, yet it leads to the spitting of $T_{\rm c}$
and significant effects on SC~\cite{machida1,machida2,machida3}.

The interplay between magnetism both with FM and  AF  and superconductivity 
is an important subject and has been discussed for long time~\cite{matsu}.
Initially the case where magnetism arises from localized moments
is considered. Thus the conduction electrons responsible for SC
is distinctively different from the magnetic sub-sysytem.
This includes chevrel compounds (RE)Rh$_4$B$_4$ and (RE)Mo$_6$S$_8$ (RE: rare earth atoms).
Magnetism affects on profound influences of SC or $H_{\rm c2}$
due to the onset of AF at $T_{\rm N}$ below which $H_{\rm c2}$ exhibits
an anomalous kink structure associated with the destruction of a part of the Fermi surface 
by AF gapping~\cite{nokura}.
In the FM case the internal FM molecular field induces Fulde-Ferrell-Larkin-Ovchinnikov state~\cite{nakanishi}
just below the Currie temperature $T_{\rm Currie}$.

Those examples of the coexistence clearly differ from the present generation of the
intertwining problem~\cite{kato,kivelson,keimer}
 in that the electrons responsible for magnetism and SC are not separable
and exhibit simultaneous roles for both orderings.
This duality of localized and itinerant electrons in the heavy Fermion materials
is essential in forming the heavy Fermion state with the enhanced electron mass.
In this case the interplay of magnetism and SC is more intricate, which is the
present situation in UTe$_2$, but as we have seen in this paper
the idea of the FM molecular field exerted from the magnetic sub-system is
quite  a useful concept in understanding various mysteries associated with the phase diagram
constructions. 
This continues to be valid and profitable to apply for other heavy Fermion SC~\cite{chris}, including
the globally or locally noncentrosymmery broken SC such as CePt$_3$Si, or CeRh$_2$As$_2$.
Those are known as the materials that AF coexists with SC, and the anomalously enhanced 
$H_{\rm c2}$ which breaks the Pauli-Clogston limit~\cite{ce}.

We admit that there are several outstanding issues to be solved in UTe$_2$
in spite of the present and previous works~\cite{machida1,machida2,machida3}.

\noindent
(1) Since according to our theory the tetra-critical point exists at $H(\parallel b)\sim$13T
as shown in Fig.~\ref{f1}(a),
the ``fourth'' second order internal phase transition is still missing.

\noindent
(2) The detailed phase diagram of HSC in $0<\theta<12^{\circ}$ is to be investigated 
because it is continuously connected to the isolated HSC around $30^{\circ}<\theta<45^{\circ}$.

\noindent
(3) The possible chiral-nonchiral transition for HSC should be checked experimentally.
The $\beta$ phase may be found.

\noindent
(4) Magnetic elastic neutron scattering experiment can probe the magnetization $M_b(H)$ component
for $0^{\circ}<\varphi<8^{\circ}$ and $0^{\circ}<\theta<45^{\circ}$ to establish our reconstructed magnetization curves as shown in 
Figs.~\ref{f2} and \ref{f3}. And also small angle neutron scattering (SANS) experiment is important to see vortices with the
spin textures for the intermediate fields of $H\parallel b$-axis.

\noindent
(5) The vortex core contains the Majorana zero energy modes spinless or spinful for HSC and LSC
respectively. Those zero Majorana modes are detected through the local density of states~\cite{ichioka}
 probed by STM, or other methods.

\section{Conclusion and summary}
Based on a nonunitary triplet pairing state, we have found that the orbital depairing limit of $H_{\rm c2}$
can be exceeded by cancelling the external field via the internal field exerted from the localized moments.
This novel mechanism for a spin triplet state allows us to analyze the $H_{\rm c2}$ phase diagrams for various field orientations 
centered along the magnetic hard $b$-axis. In particular, the record high $H_{\rm c2}\sim$ 70T occurring in between
 the $b$-axis and the $c$-axis can be understood by this orbital limit violation mechanism.
 The present work not only has identified the pairing state realized in UTe$_2$, but also
 proposed a novel mechanism for the violation of the orbital limit of $H_{\rm c2}$, which enables us to attain higher 
$H_{\rm c2}$ in a superconductor in general.

\section*{Acknowledgments}  

The author is grateful for the enlightening discussions with D. Aoki, K. Ishida, S. Kitagawa, 
Y. Shimizu, S. Kittaka, T. Sakakibara, Y. Tokunaga, Y. Haga, H. Sakai, and A. Miyake.
This work is supported by JSPS KAKENHI, No.17K05553 and No. 21K03455


\begin{thebibliography}{9}
\bibitem{dirty}
It should be noted that in a dirty superconductor,
$H_{\rm c2}$ can be arbitrarily large in principle because
the effective coherent length is given by a symbolic formula $1/\xi_{\rm eff}\propto1/l+1/\xi$
with $l$ being the transport mean free path.
\end{thebibliography}
\end{document}